\documentclass[12pt]{article}
  \newcommand{\be}{\begin{equation}}
  \newcommand{\ee}{\end{equation}}
  \newcommand{\bea}{\begin{eqnarray}}
  \newcommand{\eea}{\end{eqnarray}}

  \def\a{\alpha}

  \def\d{\delta}
  \def\e{\epsilon}
  \def\f{\phi}
  \def\g{\gamma}

  \def\p{\pi}
  
  \def\r{\rho}
  \def\s{\sigma}

  \def\O{\Omega}

\begin{document}
  \renewcommand{\theequation}{\thesection.\arabic{equation}}
  \newcommand{\eqn}[1]{eq.(\ref{#1})}
 
  \renewcommand{\section}[1]{\addtocounter{section}{1}
  \vspace{5mm} \par \noindent
    {\bf \thesection . #1}\setcounter{subsection}{0}
    \par
     \vspace{2mm} } 
  \newcommand{\sectionsub}[1]{\addtocounter{section}{1}
  \vspace{5mm} \par \noindent
    {\bf \thesection . #1}\setcounter{subsection}{0}\par}
  \renewcommand{\subsection}[1]{\addtocounter{subsection}{1}
  \vspace{2.5mm}\par\noindent {\em \thesubsection . #1}\par
   \vspace{0.5mm} }
  \renewcommand{\thebibliography}[1]{ {\vspace{5mm}\par \noindent{\bf
  References}\par \vspace{2mm}}
  \list
   {[\arabic{enumi}]}{\settowidth\labelwidth{[#1]}\leftmargin\labelwidth
   \advance\leftmargin\labelsep\addtolength{\topsep}{-4em}
   \usecounter{enumi}}
   \def\newblock{\hskip .11em plus .33em minus .07em}
   \sloppy\clubpenalty4000\widowpenalty4000
   \sfcode`\.=1000\relax \setlength{\itemsep}{0 em} }

  \begin{center}
\vspace*{3cm}
  {\bf  The longitudinal fivebrane and tachyon condensation \\
   in matrix theory} \vspace{2cm}

  Marc Massar \footnote{massar@tena4.vub.ac.be}
  and Jan Troost  \footnote{
 troost@tena4.vub.ac.be;  \,  Aspirant F.W.O.}\\
  {\em Theoretische Natuurkunde, Vrije Universiteit Brussel} \\
  {\em Pleinlaan 2, B-1050 Brussel, Belgium} \\
  \end{center}
\vspace{2cm}
  \centerline{ABSTRACT}
  \begin{quote}\small
  We study a configuration in matrix theory carying longitudinal fivebrane
 charge, i.e. a D0-D4 bound state.
 We calculate
 the one-loop effective potential between a
 D0-D4 bound state and a D0--anti-D4 bound state and 
 compare our results to
 a supergravity calculation. Next,
 we identify the tachyonic fluctuations
  in the D0-D4 and D0--anti-D4 system. We analyse classically the
  action for these tachyons and find solutions to the equations
  of motion corresponding to tachyon condensation.
  \end{quote}

\newpage
  
  \section{Introduction}
 
 Matrix theory \cite{BFSS} \cite{Su} \cite{Se} is the M-theory
 interpretation of U(N) supersymmetric quantum mechanics which has passed
 many stringent tests. The brane content of matrix theory was determined
 in \cite{BSS}. Amongst other branes, the longitudinal fivebrane was identified
 \footnote{The transverse fivebrane remained a puzzle
 \cite{L3}.}.
 Two types of representation for the
 longitudinal fivebrane were proposed. One in terms of an instanton
 gauge field, which was used in \cite{CT1} to calculate one loop
 effective potentials between the D0-D4 bound state and other
 objects in matrix theory. Another representation was proposed in
 terms of two pairs of canonical conjugate variables.  We use this
 representation to calculate  one-loop effective potentials (see
 e.g. \cite{AB} \cite{L2}  \cite{CT1}
 \cite{CT2}) between this object and a graviton, another D0-D4
 bound, and a D0--anti-D4 system. Naturally, we find agreement
 with \cite{CT1} for the cases studied there and with an extra supergravity
 calculation for the D0-D4 and D0-anti-D4 system.
 
 In \cite{Ja} a first step towards the understanding of Sen's tachyon
 condensation mechanism \cite{Sen} in matrix theory was taken, by analyzing
 the tachyon in the  D0-D2 and D0--anti-D2 system. We concentrate
 on the D0-D4 and D0--anti-D4 system. We
 identify the tachyonic
 fluctuations in the D0-D4 and D0--anti-D4 background and analyse the
 classical action for these fluctuations in the spirit of \cite{Ja}.
 We find solutions to the action
 representing condensation to a vacuum filled with D0-branes and gravitons.
 
 The first section concentrates on a discussion of the classical solution of
 matrix theory  corresponding to a D0-D4 bound state system. In the second
 section some effective potentials are calculated in detail to get
 acquainted with the representation of the longitudinal fivebrane in terms
 of canonical conjugate variables. We add a remark about the spectrum of
 the fluctuations around one longitudinal fivebrane. The next section deals
 with an analysis of the tachyonic fluctuations.
 Then we analyse possible
 solutions to the action for the tachyonic fluctuations.
 Finally, we add remarks on the
 results and  open problems.

 \section{Preliminary discussion of the classical solution}
 
 The  lagrangian of matrix theory
 is given by $U(N)$ supersymmetric quantum mechanics, namely the dimensional
 reduction of tendimensional ${\cal N}=1$ $U(N)$ Super Yang-Mills theory
 to $0+1$ dimensions. It reads
 \cite{BFSS}:
 \begin{eqnarray}
 {\cal L} &=& \frac{T_0}{2} Tr \left( (D_0 X_I)^2+ \frac{1}{2} \left[ X_{I} ,
 X_J \right]^2  + 2 \theta^{T} D_0 \theta - 2 \theta^{T} \g^I
 \left[ \theta, X_I \right] \right)
 \end{eqnarray}
 where we take $ 2 \p \a' = 1 $ and $ T_0 = \frac{\sqrt{2 \p}}{g} $.
 Furthermore we have $D_0 = \partial_t - i \left[A_0, . \right] $ and $I=1,2,
 \dots, 9$. All fields are in the adjoint of $U(N)$. The fermions are
 Majorana-Weyl.
 The equations of motion for static configurations with trivial $A_0$ and
 vanishing fermions are:
 \begin{eqnarray}
 \left[  X_I, \left[ X_I, X_J \right] \right] &=& 0.
 \end{eqnarray}
 We study especially a background configuration ($X_I =
 B_I$) corresponding to a D0-D4 bound state, or longitudinal fivebrane,
 satisfying the following
 commutation rules \cite{BSS}:
 \begin{eqnarray}
 \left[ B_1, B_2 \right] &=&  - i c \, \s_3 \otimes
 I_{\frac{N}{2} \times \frac{N}{2}} \nonumber \\
 \left[ B_3, B_4 \right] &=&  - i c \, \s_3 \otimes
 I_{\frac{N}{2} \times \frac{N}{2}},
 \end{eqnarray}
 and the other matrices and commutators zero.
 Here $ \s_3 $ is the third
 Pauli matrix and $c$ is a constant.
 We  take the infinite background matrices to be
 blockdiagonal such that this configuration solves the equations of motion.
 We will use two representations for this solution. The first one is
 in terms of two 'canonical conjugate' pairs:
 \begin{eqnarray}
 \left[ P_1, Q_1 \right] &=&  - i c  \nonumber \\
 \left[ P_2, Q_2 \right] &=&  - i c   \nonumber \\
 B_1 &=& \left( \begin{array}{cc} P_1 & 0 \\ 0 & P_1 \end{array} \right) \nonumber
\\
 B_2 &=& \left( \begin{array}{cc} Q_1 & 0 \\ 0 & -Q_1 \end{array} \right)
\nonumber \\
 B_3 &=& \left( \begin{array}{cc} P_2 & 0 \\ 0 & P_2 \end{array} \right)
\nonumber \\
 B_4 &=& \left( \begin{array}{cc} Q_2 & 0 \\ 0 & -Q_2 \end{array} \right)
 \label{bl}
 \end{eqnarray}
 This representation makes it easy to interpret the brane content
 of the configuration.
 Clearly, this solution as a whole carries no membrane charge since
 $q_2=- \frac{i}{2 \p} Tr \left[ B_I , B_J \right] = 0 $. It carries
 longitudinal fivebrane charge in the $1,2,3,4$ directions though:
 \begin{eqnarray}
 q_5=-\frac{1}{8 \p^2} \e^{IJKL} Tr \left[ B_I B_J B_K B_L \right] &
 = & N \frac{c^2}{4 \p^2}
 \end{eqnarray}
 We refer to \cite{KK} for a clear and detailed analysis of the
 charges  of the configuration, which
 yields the fact that the configuration you build in this way
 represents at least  two  D0-D4 bound states.
 That can be understood from the following reasoning.
 When we focus on the left upper block, it clearly has membrane
 charge in directions $1,2$ and $3,4$, as well as longitudinal fivebrane
 charge. It represents a D0-D4-D2-D2
 bound state. Zooming in on the right lower block we see a
 D0-D4-anti-D2-anti-D2 bound state. If we formally superimpose the
 two parts we find two D0-D4 bound states, the 2-brane charge
 cancelling out.
 
 Thinking naively, one might be worried that this superposition is
 unstable, in particular, one might expect a tachyonic off-diagonal
 mode in the background configuration, representing a string
 stretching between a D2-brane and an anti-D2-brane. We will come
 back to this point and show that there is no such tachyonic mode.
 Moreover, the configuration was shown in \cite{BSS} to preserve
 1/4 supersymmetry, as expected from  D0-D4 bound states.
 
 An alternative representation of
 the background  configuration in terms of gauge fields, discussed in detail
 in \cite{KK} will come in handy later on. It is given by:
 \begin{eqnarray}
 B^1 &=& c \left ( \begin{array}{cc} -i \partial_{x_1}  & 0  \\
                               0 &  -i \partial_{x_1}
         \end{array} \right) \nonumber \\
 B^2 &=& c \left ( \begin{array}{cc}  -i \partial_{x_2} + \frac{ x_1}{c} & 0  \\
                               0 &  -i \partial_{x_2}  - \frac{ x_1}{c}
         \end{array} \right) \nonumber \\
 B^3 &=& c \left ( \begin{array}{cc}  -i \partial_{x_3} & 0 \\
                                        0 &-i \partial_{x_3}
         \end{array} \right)  \nonumber \\
 B^4 &=& c \left ( \begin{array}{cc} -i \partial_{x_4} + \frac{ x_3}{c} & 0 \\
                     0 & -i \partial_{x_4}   - \frac{ x_3}{c}
         \end{array} \right)
 \end{eqnarray}
 Note that we introduce the same four coordinates on the two D0-D4
 bound states. This indicates our intention of treating them as a
 single object. Indeed, we will only analyse interaction potentials
 and fluctuations  where the two D0-D4 bound states move as one.
We define the left-upper  and the right-lower  part 
to be made up of
  $\frac{N_0}{2}$ D0-branes
and 
 denote the D0-brane charge density as $ \r_0 = \frac{N_0}{2 A_4}  $,
 where $A_4$ represents the (possibly infinite) area of the coinciding D4-D0
 bound states.
Then we can derive the following relation \cite{KK} :
 \begin{eqnarray}
 c^2 &=& \frac{A_4 N_4}{(2 \p)^2   N_0} \label{c}
 \end{eqnarray}
 where $N_4$ is the number of fourbranes and $N_0$ the total number of
 D0-branes in the bound state.
  \setcounter{equation}{0}
  \section{ Calculating effective potentials in matrix theory at one loop
 \footnote{
 Readers only interested in the tachyonic fluctuations can skip this
 section without much difficulty.} }
 In this section we calculate some interaction potentials between
 the D0-D4 bound state \footnote{We will
 stop mentioning that it actually consist of two D0-D4 bound states from
 now on.} and other objects explicitly. In the
 literature (e.g. \cite{AB} \cite{L2} \cite{CT1} \cite{CT2}), some
 of these potentials have already been calculated  using the
 representation in terms of an instanton background gauge field
 \cite{CT1}. But in the next section we will need a more detailed
 analysis of the fluctuations, when we identify the tachyonic ones.
 Moreover there are a few new technicalities in calculating the spectrum
 of the fluctuations when a single object is represented by
 'two-by-two' matrices, which have not been discussed in the literature yet.
 Therefore we find it useful to first redo some of the calculations in
 the literature in our representation, next to treat the new case
 of the D0-D4 and D0--anti-D4 interaction in detail.
 
 Because one object is sometimes represented by 'two-by-two'
 matrices, we need some new conventions and nomenclature, which we
 will take to be as follows. In this section, the first object will
  have extent $n_0$, the second object $N_0$. When one object
 is represented by a 'two-by-two' matrix, the submatrices will have
 half the extent of the object, e.g. $\frac{n_0}{2}$. Moreover, suppose
 we have two objects in the background represented by blockdiagonal
 'two-by-two'
 matrices. Then we
 will take the following nomenclature for the different parts of the
 coordinate matrices:
 \begin{eqnarray}
   X_I &=& \left ( \begin{array}{cccc} \mbox{block (1)} & 0 & \mbox{sector 13}
   & \mbox{sector 14} \\
                               0 & \mbox{block (2)} & \mbox{sector 23} &
                               \mbox{sector 24}  \\
                               \mbox{sector 13}^\dagger & \mbox{sector 23}^\dagger
                               & \mbox{block (3)}  & 0 \\
                               \mbox{sector 14}^\dagger & \mbox{sector
24}^\dagger
                               & 0 &
                               \mbox{block (4)}
         \end{array} \right)
 \end{eqnarray}
 The
 off-diagonal modes have been divided up into four different sectors.
 Other cases to be discussed are simpler and the nomenclature
 will be analogous in an obvious way.

 The technique to calculate the one-loop effective
 potential between two objects in matrix theory is  standard by now
 \cite{AB} \cite{L2}. To calculate the potential, we 
 determine the spectrum of the off-diagonal fluctuations corresponding
 to strings
 stretching from  one object to the other.  Their mass matrix
 is
 easily determined by expanding the action of matrix theory around
 the relevant background. This is slightly more
 involved when objects are represented by two-by-two matrices, but
 the general formulae in for instance \cite{CT1} \cite{KT} can
 easily be adapted to our case, essentially because the background
 matrices are block diagonal. We do not give the details of the
 calculation, but summarize for each case the result.
 
 In the following three
 subsections, we will discuss three different cases. For the
 first object we always take the D0-D4 bound state. For the
 second object we take respectively a graviton, a D0-D4 bound state, or a
 D0--anti-D4 bound state.  The second object will always be taken to
 be at a distance $b$ of the first object in some transverse direction
 ($"8"$)
 and it will be moving with a
 velocity $v$ relative to the first object in another transverse direction ($"9"$).
 This is incorporated by
 choosing the background matrices corresponding to these transverse
 coordinates to be:
 \begin{eqnarray}
 B_8 &=&
 \left ( \begin{array}{cc} 0_{n_0 \times n_0} & 0 \\
                                  0 & b \, I_{N_0 \times N_0}
         \end{array} \right)   \nonumber \\
 B_9 &=&
 \left ( \begin{array}{cc} 0_{n_0 \times n_0} & 0 \\
                                  0 &  v t   \, I_{N_0 \times N_0}
         \end{array} \right)
         \label{X89}
 \end{eqnarray}

 Finally,
 to make the interaction energies finite, we wrap the fourbranes on a
four-torus. This hardly influences the calculation. It is moreover convenient
 to take the four-torus to have self-dual radii $R_i = \sqrt{\a'}$. It is
 straightforward to again add in the dependence on the compactification radii
 in the final formulae. See for instance \cite{CT1}.
 
 \subsection{Interaction potential between a D0-D4 bound state system and
 a graviton}
 
 For the first case, namely the D0-D4 bound state interacting with
 a moving graviton, the non-trivial background matrices are (recall
 also the separation matrices $B_8$ and  $B_9$ given in (\ref{X89})):
 \begin{eqnarray}
 \left[ P_1, Q_1 \right] &=&  - i c \nonumber  \\ \left[ P_2, Q_2 \right]
 &=&  - i c \nonumber
 \\
  B_1 &=& \left( \begin{array}{ccc} P_1 & 0 & 0 \\ 0 & P_1 & 0 \\
  0 & 0 & 0  \end{array}
 \right) \nonumber \\
  B_2 &=& \left( \begin{array}{ccc} Q_1 &
 0 & 0 \\ 0 & -Q_1 & 0
  \\
  0 & 0 & 0  \end{array} \right) \nonumber
\\ B_3 &=& \left(
 \begin{array}{ccc} P_2 & 0 & 0 \\ 0 & P_2 & 0
  \\
  0 & 0 & 0 \end{array} \right)
\nonumber
\\ B_4 &=&
 \left( \begin{array}{ccc} Q_2 & 0 & 0 \\ 0 & -Q_2  & 0
  \\
  0 & 0 & 0  \end{array}
 \right) 
 \end{eqnarray}
For the quantum fluctuations we find
 identical spectra \footnote{Some boson contributions to the 
one loop 
effective potential are cancelled by ghost contributions. We don't
include them in the following.} in the two
 sectors of extent $\frac{n_0}{2} \times N_0$.
 We define the  hamiltonian:
  \begin{eqnarray}
 H &=& P_1^2 + Q_1^2 + P_2^2 + Q_2^2 +b^2 + v^2 t^2 \label{H},
 \end{eqnarray}
 corresponding to two non-interacting
 harmonic oscillators with frequency $c$ and a trivial extra part.
 After diagonalization, we find for the mass operators of the real
 bosons
  $ 4: H \pm 2iv;$ $ 8: H \pm 2 c ;$ $ 4 : H $ and for the  fermions
                  $8: H \pm iv; $ $ 8: H \pm  2 c \pm iv $, where
 we always state the number of fields first, and then the mass operator
 that corresponds to them. For instance $ 4: H \pm 2 c $ corresponds to 2
 fields with mass operator $ H + 2c $ and 2 fields with mass operator $H -
 2c $.
 The spectrum of these mass operators is easily determined.
 Following  \cite{AB} \cite{L2}   it is then straightforward to
 calculate the phase shift due to the interactions, and to approximate
 the phase shift at large distances $ b^2 \gg c $ and small velocities
 $v \ll b^2 $:
 \begin{eqnarray}
 \d &=& 2 \, {\cal N}  \int_0^{\infty} \frac{ds}{s}
 \frac{e^{-b^2 s}}{8 \sin{vs}  \sinh^2{c s}}  \times \nonumber \\
 & & (2+ 2 \cos{2vs}+4\cosh{2cs} \nonumber \\
 && -4 \cos{vs}-4 \cos{vs}
 \cosh{2cs}) \\
 & \approx &   (\frac{n_4 N_0 v}{2 b^2}+ \frac{v^3 n_0 N_0}{8  b^2})
 \end{eqnarray}
 We denoted the degeneracy of the energy levels by $ {\cal N}$ which in
 this case is given by:
 \begin{eqnarray}
 {\cal N} &=& c^2 \frac{n_0}{2} N_0 \label{deg} \\
 c^2 &=& \frac{n_4}{n_0}
 \end{eqnarray}
 and we have used formula (\ref{c}) at self-dual radii ($A_4=(2 \p)^2$).
 Determining the degeneracy of the spectrum  has been done for the 
 equivalent problem in the Landau model -- a charged particle in a
 magnetic field.
  The degeneracy for the Landau levels was determined in 
  \cite{AC}. Translating the formula for the degeneracy
  to our problem and carefully keeping track of normalization factors,
we find the following heuristic for
 the degeneracy in general:
 \begin{eqnarray}
 {\cal N}    &=& \mbox{Dimension fluctuation-matrix} \times \\
 & &  \mbox{Product of frequencies of the harmonic oscillators in H}.
 \end{eqnarray}
 This rule is applied
 in (\ref{deg}), $\frac{n_0 N_0}{2}$ being the dimension of the fluctuations
 and $c^2$ being the product of the harmonic oscillator frequencies in
 $H$ (\ref{H}).
 The end result for the phase shift
 matches  with the supergravity calculation in the relevant regime
 (\ref{sg1}),
 and with the result obtained in a different manner in
 \cite{CT1}. The
 phase shift starts at order $v$ because the background configuration
 preserves $ \frac{1}{4} $ supersymmetry.
  \subsection{D0-D4 and D0-D4 interaction}
 In the second case the background matrices are:
 \begin{eqnarray}
  \left[ P_1, Q_1 \right] &=&  - i c_1 \nonumber  \\ \left[ P_2, Q_2 \right]
 &=&  - i c_1 \nonumber \\
  \left[ P_3, Q_3 \right] &=&  - i c_3 \nonumber  \\ \left[ P_4, Q_4 \right]
 &=&  - i c_3 \nonumber \\
 B^1 &=& \left ( \begin{array}{cccc}  P_1 & 0 & 0 & 0 \\
                               0 &  P_1 & 0 & 0 \\
                               0 & 0 & P_3 & 0 \\
                               0 & 0 & 0 & P_3
         \end{array} \right) \nonumber  \\
 B^2 &=& \left ( \begin{array}{cccc}  Q_1 & 0 & 0 & 0 \\
                               0 &  -Q_1 & 0 & 0 \\
                               0 & 0 & Q_3 & 0 \\
                               0 & 0 & 0 & -Q_3
         \end{array} \right)  \nonumber \\
 B^3 &=& \left ( \begin{array}{cccc}  P_2 & 0 & 0 & 0 \\
                               0 &  P_2 & 0 & 0 \\
                               0 & 0 & P_4 & 0 \\
                               0 & 0 & 0 & P_4
         \end{array} \right) \nonumber \\
 B^4 &=& \left ( \begin{array}{cccc}  Q_2 & 0 & 0 & 0 \\
                               0 &  -Q_2 & 0 & 0 \\
                               0 & 0 & Q_4 & 0 \\
                               0 & 0 & 0 & -Q_4
         \end{array} \right) 
 \end{eqnarray}
  Here we find four sectors of extent $\frac{n_0}{2} \times
 \frac{N_0}{2}$, two by two identical, namely sector $(13)=(24)$
 and sector $(23)=(14)$. We define the two relevant hamiltonians
  \begin{eqnarray}
 H^{(13)} &=& (P_1+P_3)^2 + (Q_1-Q_3)^2
          + (P_2+P_4)^2 + (Q_2-Q_4)^2 +b^2 + v^2 t^2  \nonumber \\
 H^{(23)} &=& (P_1+P_3)^2 + (Q_1+Q_3)^2
          + (P_2+P_4)^2 + (Q_2+Q_4)^2 +b^2 + v^2 t^2. \nonumber
 \end{eqnarray}
 Each describes a system of two
 decoupled harmonic oscillators.
 The diagonalized mass operators are: in sector (13)
 for the bosons  $4: H \pm 2iv ;$ $ 8: H \pm 2 (c_1-c_3) ; $ $
 2 : H $ and for the fermions $8: H \pm iv ; $ $ 8 : H \pm iv \pm  2
 (c_1-c_3)$.
 In sector
           (23) they read for the bosons  $4: H \pm 2iv; $ $8 : H \pm 2 (c_1+c_3) ;$ $
 4 : H $ and for the fermions $8: H \pm iv ;$ $ 8 : H \pm iv \pm  2(c_1+c_3) $.
 The spectrum is again easily determined, and the phase shift now
 gets two different contributions:
 \begin{eqnarray}
 \d_{(13)+(24)} &=& 2 \, {\cal N}_{13}  \int_0^{\infty} \frac{ds}{s}
 \frac{e^{-b^2 s}}{8 \sin{vs}  \sinh^2{(c_1-c_3) s}} \times \nonumber \\ &
 & (2+ 2 \cos{2vs}+4\cosh{2(c_1-c_3)s} \nonumber \\ & & -4
 \cos{vs}- 4 \cos{vs} \cosh{2(c_1-c_3)s})
  \\
 & \approx & {\cal N}_{13}  (\frac{v}{ b^2}+ \frac{v^3 }{4
 (c_1-c_3)^2 b^2}) \\ \d_{(23)+(14)} &=& 2 \, {\cal N}_{23}
 \int_0^{\infty} \frac{ds}{s} \frac{e^{-b^2 s} }{8 \sin{vs}
 \sinh^2{(c_1+c_3) s}} \nonumber \times \\ & & (2+ 2
 \cos{2vs}+4\cosh{2(c_1+c_3)s} \nonumber \\ & & -4 \cos{vs}- 4
 \cos{vs} \cosh{2(c_1+c_3)s})
 \\
 & \approx & {\cal N}_{23}  (\frac{v}{ b^2}+ \frac{v^3 }{4 (c_1+c_3)^2 b^2})
 \end{eqnarray}
 giving a total phase shift
 \begin{eqnarray}
 \d & \approx &   \frac{ (n_0 N_4+N_0 n_4) v}{2 b^2}+ \frac{n_0 N_0 v^3 }{8  b^2}
\label{d4d4}
 \end{eqnarray}
 where we have used the following formulae:
 \begin{eqnarray}
 {\cal N}_{13} &=& (c_1-c_3)^2 \frac{n_0}{2} \frac{N_0}{2} \\
 {\cal N}_{23} &=& (c_1+c_3)^2 \frac{n_0}{2} \frac{N_0}{2} \\
 c_1^2 &=& \frac{n_4}{n_0}                                 \\
 c_3^2 &=& \frac{N_4}{N_0} .
 \end{eqnarray}
 The fact that the phase shift starts at order $v$ is due to the fact
 that the background configuration preserves 1/4 supersymmetry. The
 endresult matches with the supergravity calculation (\ref{sg2}) \cite{CT1}.
  \subsection{D0-D4 and D0-anti-D4 interaction}
 In the third case the background matrices are:
 \begin{eqnarray}
  \left[ P_1, Q_1 \right] &=&  - i c_1 \nonumber  \\ \left[ P_2, Q_2 \right]
 &=&  - i c_1 \nonumber \\
  \left[ P_3, Q_3 \right] &=&  - i c_3  \nonumber \\ \left[ P_4, Q_4 \right]
 &=&  - i c_3 \nonumber \\
 B^1 &=& \left ( \begin{array}{cccc}  P_1 & 0 & 0 & 0 \\
                               0 &  P_1 & 0 & 0 \\
                               0 & 0 & -P_3 & 0 \\
                               0 & 0 & 0 & -P_3
         \end{array} \right)   \nonumber \\
 B^2 &=& \left ( \begin{array}{cccc}  Q_1 & 0 & 0 & 0 \\
                               0 &  -Q_1 & 0 & 0 \\
                               0 & 0 & Q_3 & 0 \\
                               0 & 0 & 0 & -Q_3
         \end{array} \right)  \nonumber \\
 B^3 &=& \left ( \begin{array}{cccc}  P_2 & 0 & 0 & 0 \\
                               0 &  P_2 & 0 & 0 \\
                               0 & 0 & P_4 & 0 \\
                               0 & 0 & 0 & P_4
         \end{array} \right) \nonumber \\
 B^4 &=& \left ( \begin{array}{cccc}  Q_2 & 0 & 0 & 0 \\
                               0 &  -Q_2 & 0 & 0 \\
                               0 & 0 & Q_4 & 0 \\
                               0 & 0 & 0 & -Q_4
         \end{array} \right) 
 \end{eqnarray}
 Note the partial sign change in the first background matrix,
 turning the second object into a D0--anti-D4 bound state. We find
 four sectors of extent $\frac{n_0}{2} \times \frac{N_0}{2}$, all
 with identical spectra, when we ignore the origin in terms of the
 different coordinates \footnote{We can do so for calculating the
 effective potential, but in  section 5 we need the precise
 origin of the tachyonic modes in terms of the coordinate
 matrices. We return there to this point.}. The relevant hamiltonian is:
  \begin{eqnarray}
 H^{(13)} &=& (P_1-P_3)^2 + (Q_1-Q_3)^2
          + (P_2+P_4)^2 + (Q_2-Q_4)^2 +b^2 + v^2 t^2, \nonumber
 \end{eqnarray}
 corresponding to a system of two harmonic oscillators. We will
 always suppose that $ c_1 - c_3  $ is  positive,
 the other case being fully equivalent. The
 mass operators are for each sector
  for the bosons  $4: H \pm 2iv ;$ $ 4 : H \pm 2 (c_1+c_3); $ $ 4 :
 H \pm 2 (c_1-c_3) ;$ $ 4 : H $ and for the fermions $ 16: H \pm iv  \pm
 (c_1+c_3) \pm (c_1-c_3) $. The potential is then :
 \begin{eqnarray}
 {\cal V} &=& \frac{2}{\sqrt{\p}} \, {\cal N}  \int_0^{\infty} \frac{ds}{s} 
\frac{e^{-b^2 s}}{4 s^{1/2}
 \sinh{(c_1-c_3) s} \sinh{(c_1+c_3) s}} \times  \nonumber \\
 & & (2+
 2 \cos{2vs}+2\cosh{2(c_1-c_3)s}+2\cosh{2(c_1+c_3)s} \nonumber \\
 & &-8 \cos{vs}
 \cosh{(c_1+c_3)s} \cosh{(c_1-c_3)s})\\
 & \approx & \frac{ n_4 N_4}{ b^3}+
 \frac{(n_0 N_4+N_0 n_4) v^2}{4 b^3}+ \frac{n_0 N_0 v^4 }{16 b^3}
 \end{eqnarray}
 Compared to the previous case (\ref{d4d4}), there
is an extra interaction between the D4-brane and the anti-D4-brane.
 The interaction potential
is non-trivial at zero velocity and the
background fully breaks supersymmetry.
 The end
 result is reproduced by our
 supergravity calculation (\ref{sg3}) in the appendix.
 Clearly, the formula for the potential breaks down at small distances
$b^2 \le 2 c_3$. Then there is a tachyon in the spectrum of the bosons since
the lowest energy mode has mass:
$E=(c_1-c_3) + (c_1+c_3)+b^2 - 2 (c_1+c_3)= b^2- 2 c_3$. We will treat the system at 
short distances in   section 5.
 
 \subsection{Summary}
 The conclusions we draw from these calculations are the following.
 At the level  we are probing the system, the representation
 of the D0-D4 system that we use is equivalent to the
 instanton gauge field representation used in \cite{CT1}.
 We found full
 agreement when we compared the long range one loop potentials with
 supergravity results, also for the case of the D0-D4 and the
 D0--anti-D4 system, as expected. Moreover, we showed
 that it makes perfect sense to divide the off-diagonal modes into
 different sectors and treat them separately,
 which will be important in the second part of our paper.
 
\setcounter{equation}{0}
 \section{Remark on the fluctuations around one longitudinal fivebrane}
 We refer to \cite{KK} for an analysis of the effective
 action for the fluctuations around the  D0-D4 bound state system, but we 
 add a remark that fits well into the context of our paper.
  As we mentioned in section 2, you might expect a tachyonic
 off-diagonal mode in the coordinate matrices spanning the
 fivebranes (\ref{bl}), since they could correspond to strings stretching from
 a D2-brane to an anti-D2-brane. That this does not happen is shown
 by a small calculation. The relevant mass matrix for these modes
 is, for instance for the fluctuations in the
 coordinates $X_1$ and $X_2$:
 \begin{eqnarray}
 M_{12} &=& \left ( \begin{array}{cc} H & -2 i c \\
                                  2 i c  & H
         \end{array} \right)
 \end{eqnarray}
 where \begin{eqnarray} H &=& P_1^2 + Q_1^2 + P_2^2 + Q_2^2.
 \end{eqnarray}
 Diagonalizing the mass matrix and determing the spectrum yields
 two kinds of fluctuations with the following energies:
 \begin{eqnarray} E &=& c (2n+1) + c (2m+1) + 2c \\ E' &=& c (2
 n'+1) + c(2 m'+1) -2c
 \end{eqnarray}
 Note that for the last kind of fluctuation, we find a  massless
  mode, and not a tachyonic one. This is due to the quantummechanical
 zero point
 energies coming from the object spanning in the 1,2 as well as the
 3,4 direction. For a membrane--anti-membrane system this mode
 would be tachyonic \cite{AB} \cite{Ja}.
\setcounter{equation}{0}
 \section{The action for the tachyonic fluctuations}
 \subsection{The tachyonic fluctuations}
 From now on, we will consider the D0-D4 system and the D0--anti-D4  system to
 lie on top of each other, so we put the background matrices $B_8$ and $B_9$
(\ref{X89})
 to zero.  Then, when we compute  the mass matrix for the
 fluctuations in the coordinate matrices $X_1$ and
  $X_2$, we find the following
 matrix for sector 13:
 \begin{eqnarray}
 M^{(13)}_{12} &=& \left ( \begin{array}{cc} H^{(13)} & -2 i
 (c_1+c_3)
 \\
                                  2 i (c_1+c_3)  & H^{(13)}
         \end{array} \right)
 \end{eqnarray}
 where \begin{eqnarray} H^{(13)} &=& (P_1-P_3)^2 + (Q_1-Q_3)^2 +
 (P_2+P_4)^2 + (Q_2-Q_4)^2 .
 \end{eqnarray}
 We  diagonalize the mass matrix and determine the spectrum
 for the diagonal fluctuations
  \footnote{Recall that we chose $c_1 \ge c_3$.}:
 \begin{eqnarray} E &=& (c_1+c_3) (2n+1) + (c_1-c_3) (2m+1) + 2(c_1+c_3) \label{spec}
                 \\ E' &=& (c_1+c_3) (2n'+1) + (c_1-c_3) (2 m'+1) -2
                 (c_1+c_3)
 \end{eqnarray}
 From the second line, we find a tachyonic mode, as expected, with mass $-2 c_3$.
  Note that for  $c_1  < 2 c_3 $ you find several
 tachyonic modes.
 When you follow the simple diagonalization
 procedure in detail, you find that the tachyonic fluctuation
\footnote{By abuse of language, we take 
'tachyonic fluctuation'  to mean that the field includes a tachyonic
mode.} is:
 \begin{eqnarray}
 \f &=& \frac{y^{(13)}_2 - i y^{(13)}_1}{\sqrt{2}}
 \end{eqnarray}
 where $y^{(mn)}_I$ denotes the fluctuation in  sector $(mn)$ and
 coordinate matrix $X_I$. The fluctuation
 \begin{eqnarray}
 \bar{\f} &=& \frac{y^{(13)}_2 + i y^{(13)}_1}{\sqrt{2}}
 \label{if}
 \end{eqnarray} corresponds to  (\ref{spec}) and is
 never tachyonic. In the other
 sectors the computation goes analogously for a total of
  four tachyonic fields that
 correspond to strings stretching between the two D4 branes and the
 two anti-D4 branes (in the presence of the D0-branes). They are
  given by:
 \begin{eqnarray}
 \f &=& \frac{y^{(13)}_2 - i y^{(13)}_1}{\sqrt{2}} \nonumber \\ \f' &=&
 \frac{y^{(24)}_2 + i y^{(24)}_1}{\sqrt{2}} \nonumber \\ \chi &=&
 \frac{y^{(14)}_4 - i y^{(14)}_3}{\sqrt{2}} \nonumber \\ \chi' &=&
 \frac{y^{(23)}_4 + i y^{(23)}_3}{\sqrt{2}} 
 \end{eqnarray}
 
 \subsection{The action}
 Next we turn to the analysis of the action for the tachyonic
 fluctuations in the spirit of \cite{Ja}. We expand the classical action around the
 D0-D4 and D0--anti-D4 background, only keeping track of the tachyonic
 fluctuations and the gauge fields of the unbroken gauge group $U(1)^4$
 under which the tachyons are charged. All fields we believe to be
 irrelevant, we put to zero, for instance (\ref{if}).
 For simplicity,
 we take the number of D0-D4 bound states  and D0-anti-D4 bound states
 to be equal, i.e. $ c_1 = c_3 = c$.
 Now the second representation introduced in section 2
 comes in handy.
 Under the preceding assumptions, the coordinate matrices are given by:
 \begin{eqnarray}
 X^1 &=& c \left ( \begin{array}{cccc}  - i \partial_{x_1} +
 A^{(1)}_{x_1}+a^{(1)}_{x_1} & 0 & i \frac{\f}{\sqrt{2} c} & 0 \\
                               0 &  - i \partial_{x_1} + A^{(2)}_{x_1}+a^{(2)}_{x_1} & 0 &
                     -i\frac{\f'}{\sqrt{2}c} \\
                               -i \frac{\f^{\ast}}{\sqrt{2}c} & 0 & - i \partial_{y_1} + A^{(3)}_{y_1}+a^{(3)}_{y_1} & 0 \\
                               0 & i \frac{{\f '}^{\ast}}{\sqrt{2}c} & 0 &
                               - i \partial_{y_1} + A^{(4)}_{y_1}+a^{(4)}_{y_1}
         \end{array} \right) \nonumber  \\
 X^2 &=& c \left ( \begin{array}{cccc}  - i \partial_{x_2} +
 A^{(1)}_{x_2}+a^{(1)}_{x_2} & 0 &  \frac{\f}{\sqrt{2}c} & 0 \\
                               0 &  - i \partial_{x_2} + A^{(2)}_{x_2}+a^{(2)}_{x_2} & 0 &
                                 \frac{\f'}{\sqrt{2}c} \\
                               \frac{\f^{\ast}}{\sqrt{2}c} & 0 & - i \partial_{y_2} + A^{(3)}_{y_2}+a^{(3)}_{y_2} & 0 \\
                               0 &  \frac{{\f '}^{\ast}}{\sqrt{2}c} & 0 & - i \partial_{y_2} + A^{(4)}_{y_2}+a^{(4)}_{y_2})
         \end{array} \right) \nonumber \\
 X^3 &=& c \left ( \begin{array}{cccc}  - i \partial_{x_3} +
 A^{(1)}_{x_3}+a^{(1)}_{x_3} & 0 & 0 &  i \frac{\chi}{\sqrt{2}c} \\
                               0 &  - i \partial_{x_3} + A^{(2)}_{x_3}+a^{(2)}_{x_3} &
                               -i\frac{\chi'}{\sqrt{2}c} & 0 \\
                               0 & i\frac{{\chi '}^{\ast}}{\sqrt{2}c} & - i \partial_{y_3} + A^{(3)}_{y_3}+a^{(3)}_{y_3} & 0 \\
                               -i\frac{\chi^{\ast}}{\sqrt{2}c} & 0 & 0 & - i \partial_{y_3} + A^{(4)}_{y_3}+a^{(4)}_{y_3}
         \end{array} \right)  \nonumber \\
 X^4 &=& c \left ( \begin{array}{cccc}  - i \partial_{x_4} +
 A^{(1)}_{x_4}+a^{(1)}_{x_4} & 0 & 0 & \frac{\chi}{\sqrt{2}c} \\
                               0 &  - i \partial_{x_4} + A^{(2)}_{x_4}+a^{(2)}_{x_4} &
                               \frac{\chi'}{\sqrt{2}c} & 0 \\
                               0 & \frac{{\chi '}^{\ast}}{\sqrt{2}c} & - i \partial_{y_4} + A^{(3)}_{y_4}+a^{(3)}_{y_4} & 0 \\
                              \frac{\chi^{\ast}}{\sqrt{2}c} & 0 & 0 &
                               - i \partial_{y_4} + A^{(4)}_{y_4}+a^{(4)}_{y_4}
         \end{array} \right)           \nonumber
 \end{eqnarray}
where $A$ is the
 background gauge field and $a$ the gauge field fluctuation.
 The background is invariant under $U(1)^4$, each $U(1)$ has its own
 upper index. We choose the background gauge fields 
 such that the appropriate commutation relations
 between the background matrices are satisfied:
 \begin{eqnarray}
 A^{(1)}_{x_2} = - A^{(2)}_{x_2} &=& \frac{x_1}{c} \nonumber \\
 A^{(1)}_{x_4} = - A^{(2)}_{x_4} &=& \frac{x_3}{c} \nonumber  \\
 A^{(3)}_{y_2} = -A^{(4)}_{y_2}  &=& - \frac{y_1}{c} \nonumber  \\
 A^{(3)}_{y_4} = - A^{(4)}_{y_4} &=& \frac{y_3}{c}   \label{bg},
 \end{eqnarray}
 and the rest zero.
 Each tachyonic mode is charged under two of the abelian gauge
  symmetries, with opposite charges, as can easily be seen by looking at
  the transformation properties of the full coordinate matrix.
 
 To represent the action in terms of an integral over the worldvolume of the
 branes, we use the rules of \cite{BSS}, improved  in \cite{KK} and elaborated
 upon in \cite{Ja}. We rally some of the technical details to  appendix B.
 The following definitions come in
 handy in writing down the endresult.
 The non-center-of-mass coordinates are:
 \begin{eqnarray}
 u_i &=& \frac{ x_i+y_i}{2}.
 \end{eqnarray}
 Covariant
 derivatives and field strengths are defined as (Upper indices 
  label the gauge symmetries, lower indices $w_i =(x_i,y_i)$
  label coordinates.) :
 \begin{eqnarray}
 \nabla^{(\pm m)}_{w_i} &=& \partial_{w_i} \pm
 iA^{(m)}_{w_i} \pm i a^{(m)}_{w_i}
 \nonumber \\
 F^{(m)}_{w_i w_j} &=& i \left[ \nabla^{(m)}_{w_i} ,
 \nabla^{(m)}_{w_j} \right] \nonumber \\
 \nabla^{(m,\pm n)}_{u_i} &=& \nabla^{(m)}_{x_i} + \nabla^{(\pm n)}_{y_i} 
\nonumber \\
 F^{(m, \pm n)}_{u_i u_j} &=& i \left[ \nabla^{(m, \pm n)}_{u_i} ,
 \nabla^{(m, \pm n)}_{u_j} \right]
\nonumber \\
 &=& F^{(m)}_{x_i x_j} \pm F^{(n)}_{y_i y_j} 
 \label{def}
 \end{eqnarray}
 By a small $f$ we will denote the field strength $F$
 without the background gauge fields contribution.
 The relevant part of the
 action for the fluctuations that we consider
   is then given by $S=\int d^4 u \, {\cal L} $, 
and the lagrangian by (up to an overall factor)  :
 \begin{eqnarray}
 -{\cal L} &=& 
  ( \frac{c^2}{2} f^{(1,-3)}_{u_1 u_2} - c + |\f|^2)^2+
   ( \frac{c^2}{2} f^{(1,-4)}_{u_3 u_4} - c + |\chi|^2)^2
    \nonumber \\
 & & + ( \frac{c^2}{2} f^{(2,-4)}_{u_1 u_2} + c - |\f'|^2)^2+
   ( \frac{c^2}{2} f^{(2,-3)}_{u_3 u_4} + c - |\chi'|^2)^2
  \nonumber  \\
   & & 
 +\frac{c^2}{2} ( |(\nabla^{(1,-3)}_{u_2}+ i
 \nabla^{(1,-3)}_{u_1}) \f|^2  + 2 |\nabla_{u_3}^{(1,-3)} \f|^2+ 2 |\nabla_{u_4}^{(1,-3)} \f|^2 
\nonumber \\
 & & 
 +|(\nabla^{(1,-4)}_{u_4}+ i
 \nabla^{(1,-4)}_{u_3}) \chi|^2  + 2 |\nabla_{u_1}^{(1,-4)} \chi|^2 + 2 |\nabla_{u_2}^{(1,-4)} \chi|^2  
\nonumber \\
  & & 
 +|(\nabla^{(2,-4)}_{u_2}- i
 \nabla^{(2,-4)}_{u_1}) \f'|^2  + 2 |\nabla_{u_3}^{(2,-4)} \f'|^2+ 2 |\nabla_{u_4}^{(2,-4)} \f'|^2
\nonumber \\
 & & 
 +|(\nabla^{(2,-3)}_{u_4}- i
 \nabla^{(2,-3)}_{u_3}) \chi'|^2  + 2 |\nabla_{u_1}^{(2,-3)} \chi'|^2
 + 2 |\nabla_{u_2}^{(2,-3)} \chi'|^2 )  \nonumber \\
 & & + \frac{c^4}{4} (f^{(1,3)^2}_{u_1 u_3}+
 f^{(1,-3)^2}_{u_1 u_3}  + f^{(2,4)^2}_{u_1 u_3}+
 f^{(2,-4)^2}_{u_1 u_3} +
 f^{(1,3)^2}_{u_1 u_3}+
 f^{(1,-3)^2}_{u_2 u_3}  + f^{(2,4)^2}_{u_2 u_3}+
 f^{(2,-4)^2}_{u_2 u_3}    \nonumber \\
 & &  + f^{(1,3)^2}_{u_1 u_4}+
 f^{(1,-3)^2}_{u_1 u_4}  + f^{(2,4)^2}_{u_1 u_3}+
 f^{(2,-4)^2}_{u_1 u_4} +
f^{(1,3)^2}_{u_2 u_4}+
 f^{(1,-3)^2}_{u_2 u_4}  + f^{(2,4)^2}_{u_2 u_4}+
 f^{(2,-4)^2}_{u_2 u_4}     \nonumber \\
 & &  + f^{(1,3)^2}_{u_1 u_2}+
 f^{(2,4)^2}_{u_1 u_2}  +
 f^{(1,4)^2}_{u_3 u_4} +              f^{(2,3)^2}_{u_3 u_4}  )  \nonumber \\
 & & + |(\phi {\chi '}^{\ast } - \chi {\phi '}^{\ast })|^2  +
  |(\phi \chi^{\ast } - \chi' {\phi '}^{\ast })|^2 
 \label{action}
 \end{eqnarray}
where all fields only depend on the non-center-of-mass coordinates.
 Note that it is the Lagrangian you expect, with the usual kinetic terms
 for the gauge fields, the appropriate covariant derivatives hitting the
 tachyons and a Higgs potential for the tachyons. There are some 
 interactions between the tachyons and the gauge fields  \cite{Ja},
 and an interaction potential between the different tachyons.

 \subsection{Boundary conditions}
 The background gauge fields corresponding to the diagonal U(1)'s
 (\ref{bg}) can be rewritten as follows:
 \begin{eqnarray}
 {\cal A}_{u_1} &=& 0 \nonumber \\
 {\cal A}_{u_2} &=& \left ( \begin{array}{cccc}  A_{u_2}^{(1)} & 0 & 0 & 0 \\
                               0 &   A_{u_2}^{(2)}& 0 & 0 \\
                               0 & 0 &  A_{u_2}^{(3)} & 0 \\
                               0 & 0 & 0 &  A_{u_2}^{(4)}
         \end{array} \right) \nonumber   \\
         &=&     \frac{u_1}{c} \left     ( \begin{array}{cccc}  1 & 0 & 0 & 0 \\
                               0 &   -1  & 0 & 0 \\
                               0 & 0 &  -1 & 0 \\
                               0 & 0 & 0 &  1
         \end{array} \right) \nonumber   \\
 {\cal A}_{u_3} &=& 0  \nonumber \\
 {\cal A}_{u_4} &=& \frac{u_3}{c}   \left   ( \begin{array}{cccc}  1 & 0 & 0 & 0 \\
                               0 &   -1  & 0 & 0 \\
                               0 & 0 &  1 & 0 \\
                               0 & 0 & 0 &  -1
         \end{array} \right)
 \end{eqnarray}
 The non-zero background gauge fields appearing in the covariant
 derivatives in the kinetic terms for the tachyons are:
 \begin{eqnarray}
 {\cal A}^{(1,-3)}_{u_2} &=& \frac{2 u_1}{c} \nonumber \\
 {\cal A}^{(2,-4)}_{u_2} &=& -\frac{2 u_1}{c} \nonumber \\
 {\cal A}^{(2,-3)}_{u_4} &=& -\frac{2 u_3}{c} \nonumber \\
 {\cal A}^{(1,-4)}_{u_4} &=& \frac{2 u_3}{c}
 \end{eqnarray}
 Taking the  background gauge fields to live on a four-torus
with radii $R_{u_i}$, they satisfy 't Hooft's twisted boundary conditions  
\cite{H}. They read in direction $u_1$  :
 \begin{eqnarray}
 {\cal A}_{u_i} (R_{u_1},u_2,u_3,u_4) &=& - i \O_{u_1} \partial_{u_i}
 \O_{u_1}^{-1} + \O_{u_1} {\cal A}_{u_i} (0,u_2,u_3,u_4)  \O_{u_1}^{-1}
 \end{eqnarray}
 and analogous for the other directions,
where $ \O_{u_i} $
are the transition functions. The transition functions can be
 choosen to be:
 \begin{eqnarray}
 \O_{u_1} &=& \exp{[ -i u_2 \frac{R_{u_1}}{c}
 \left( \begin{array}{cccc}  1 & 0 & 0 & 0 \\
                               0 &   -1  & 0 & 0 \\
                               0 & 0 &  -1 & 0 \\
                               0 & 0 & 0 &  1
         \end{array} \right)]} \nonumber \\
 \O_{u_2} &=&  1      \nonumber  \\
 \O_{u_3} &=& \exp{[-i u_4 \frac{R_{u_3}}{c}
 \left( \begin{array}{cccc}  1 & 0 & 0 & 0 \\
                               0 &   -1  & 0 & 0 \\
                               0 & 0 &  1 & 0 \\
                               0 & 0 & 0 &  -1
         \end{array} \right)]} \nonumber \\
 \O_{u_4} &=&   1       \label{tm}
 \end{eqnarray}
 These boundary conditions are due to the presence of the \em background
 \em field, i.e. due to the  magnetic field made up of the
 D0-branes, representing the background objects.
 For the full  background matrix this implies:
 \begin{eqnarray}
 B_{I} (R_{u_1},u_2,u_3,u_4) &=& \O_{u_1} B_I (0,u_2,u_3,u_4) \O_{u_1}^{-1}
 \label{bbc}
 \end{eqnarray}
 and analogously for the other directions.

 The boundary conditions for the tachyons that are trivial \em with respect
 to the background \em can be read of from (\ref{bbc}):
 \begin{eqnarray}
\f (u_1=R_1) &=& \f(u_1=0) e^{-2 i u_2 R_1 /c} \nonumber \\
 \f' (u_1=R_1) &=& \f'(u_1=0) e^{2 i u_2  R_1 /c} \nonumber \\
 \chi (u_3=R_3) &=& \chi(u_3=0) e^{-2 i u_4 R_3 / c} \nonumber \\
 \chi' (u_3=R_3) &=& \chi'(u_3=0) e^{2 i u_4  R_3 / c} 
     \label{tbbc}
 \end{eqnarray}
 and the other background boundary conditions are trivial.
 
\setcounter{equation}{0}
 \section{A solution to the equations of motion}
 
 First we look for a solution to the equations of motion where the total 
 Lagrangian (\ref{action}) vanishes and the background boundary
conditions are satisfied.
 We make the following
 ansatz:
 \begin{eqnarray}
\phi &=& {\phi '} ^{ \ast}  (u_1, u_2) \nonumber \\
 \chi &=& {\chi '}^{\ast} (u_3,u_4)
  \label{ans}.
 \end{eqnarray}
 Then we find we can take:
 \begin{eqnarray}
 & & a^{(1)}_{u_{1,2}} = - a^{(2)}_{u_{1,2}}  = - a^{(3)}_{u_{1,2}} 
 =  a^{(4)}_{u_{1,2}}  \nonumber \\
& & a^{(1)}_{u_{3,4}} = - a^{(2)}_{u_{3,4}}  =  a^{(3)}_{u_{3,4}} 
 = - a^{(4)}_{u_{3,4}}. \label{gs}
 \end{eqnarray}
 The remaining non-trivial equations are:
 \begin{eqnarray}
 \frac{c^2}{2} f^{(1,-3)}_{u_1 u_2} - c + |\f|^2 &=& 0 \nonumber \\
 (\nabla^{(1,-3)}_{u_2}+ i
 \nabla^{(1,-3)}_{u_1}) \f &=& 0 \nonumber \\
 \frac{c^2}{2} f^{(1,-4)}_{u_3 u_4} - c + |\chi|^2 &=& 0  \nonumber \\
 (\nabla^{(1,-4)}_{u_4}+ i
 \nabla^{(1,-4)}_{u_3}) \chi &=& 0
 \end{eqnarray}
 Under the assumption (\ref{ans}), we get two copies of the
  Bogomolny equations. These have been studied in the context of Chern-Simons
 theory in detail \cite{JP} \cite{O} and we only summarize some main features.
 We can find magnetic soliton 
solutions to these equations with the
 background boundary conditions (\ref{tbbc}). Since the spatial
 worldvolume of the D4-brane
is fourdimensional, and the tachyons have non-trivial winding number
around a circle at infinity,
the magnetic solitons are twodimensional.
 The boundary conditions are
treated in detail in \cite{Ja}.
Using the solutions, we calculate the D0-brane charge from the 
 worldvolume action of the D4-branes:
 \begin{eqnarray}
 N &=& \frac{1}{8 \p^2} \int d^4 u \left(F^{(1)} F^{(1)} + F^{(2)} F^{(2)} -
 F^{(3)} F^{(3)} -F^{(4)} F^{(4)} \right) \\
&=& \frac{1}{4 \p^2} \int d^4 u F^{(1,-3)}_{u_1 u_2} F^{(1,-4)}_{u_3 u_4}
\nonumber \\
&=& \frac{A_4}{c^2 \p^2},
 \end{eqnarray} which is the original D0-brane charge.
The D0 charge is concentrated at the intersections of the
 orthogonal twodimensional solitons. Moreover, from (\ref{gs})
we find that the D2-brane charge cancels. This is consistent with the fact
that we find, from the commutators
(\ref{com}) and (\ref{com2}), and the supersymmetry variations 
 \begin{eqnarray}
 \d \theta &=& \frac{1}{2} \left( D_0 X^I \g_I +\frac{1}{2} \left[X^I,X^J
\right]
\g_{IJ} \right) \e + \e'
 \label{susy}
\end{eqnarray}
 that the tachyon condensation restores all dynamical
supersymmetry.
 We conclude that the end products after tachyon condensation 
are the original D0-branes, and extra gravitons as argued in \cite{Ja}.

\setcounter{equation}{0}
\section{Remarks and conclusion}
In the previous section, we considered tachyon condensation where the 
tachyons had trivial boundary conditions relative to the background.
We can consider more general possibilities, where the tachyons satisfy 
different boundary conditions. In the case of a
 membrane--anti-membrane configuration, this amounts to the following.
 By choosing the topological sector of the tachyon
on the D2-brane anti-D2-brane to be non-trivial, 
one can add or subtract D0-brane charge.
After condensation, this gives an arbitrary
number of D0-branes.  
 Technically, this is a trivial extension of \cite{Ja}.
In particular, the approximate solution to the equations of
 motion in \cite{Ja} remains practically unchanged.
In the case of the D0-D4 and D0-anti-D4, we have more possibilities.
For instance, 
by changing the topological sectors of the four tachyons simultaneously, we
can modify the amount of D0-brane charge in the end product in a fairly
obvious
manner (keeping the condition (\ref{ans})). It is clear that for a more
general choice of topological sectors, the end product will have D2-brane
charge.  It would be interesting to study
such condensation in detail. 
 
In this paper we have studied the interactions between a D0-D4 bound state and 
a D0-anti-D4 bound state in matrix theory. First, we calculated the interaction
potential at large distances and succesfully compared the result to an
equivalent supergravity calculation.  Next, we looked at a coinciding D0-D4 and
D0--anti-D4 bound state system and identified the tachyonic fluctuations.
We derived
 the classical action for these tachyonic fluctuations and found
solutions to the equations of motion corresponding to tachyon condensation to
D0-branes.

  \vspace{1cm}
 
  \noindent {\bf Acknowledgments}: We would like to thank  Richard
  Corrado,  Ben Craps, Shiraz Minwalla,
  Frederik Roose, Alex Sevrin and Walter Troost for useful discussions. This work
was supported in part by the European Commission TMR programme
ERBFMRX-CT96-0045 in which the authors are associated to K.U.Leuven.
 
  \newpage
  \appendix
  \noindent
\setcounter{equation}{0}
   {\bf APPENDIX}
  \section{The probe-background  calculation}
 The standard technique for calculating the interaction potential
 (or corresponding phase shift) between
 two objects from the Born-Infeld action
 and supergravity approach is the following.
 You treat one object as the background and take the
 corresponding solution of the supergravity equations of motion. Next, you
 consider the worldvolume action of the other object in this background and
 calculate the potential it feels due to the background. This has been done for many situations in the literature (see for
 instance \cite{CT1} \cite{CT2}). We  state the results of these calculations for comparison
 with the results obtained for matrix theory in the body of the paper.
 We take the conventions of \cite{CT1} ($2 \p \a'=1$)
 and we work at self-dual radius ($R_i = \sqrt{\a'}$) for the compactified
 directions. We moreover approximate the potential at large distances and
 small relative velocities between the two objects. For the interaction between
 a D0-brane bound state
 and a D0-D4 bound state we find the following phase shift $ \d $ \cite{CT1}:
 \begin{eqnarray}
  \d & \approx &   \frac{1}{2 b^2} N_0 \left[ n_4 v +\frac{1}{4} (n_0 +2 n_4) v^3 \right] +
  O (\frac{1}{b^5},v^5).                    \label{sg1}
 \end{eqnarray}
 For the interaction between two D0-D4 bound states, we find \cite{CT1}:
 \begin{eqnarray}
  \d & \approx &   \frac{1}{2 b^2} \left[(n_0 N_4 + N_0 n_4)
  v +\frac{1}{4}( n_0 N_0 + n_4 N_4 +2 n_0 N_4 + 2 n_4 N_0) v^3 \right]
\nonumber \\
& & 
+O(\frac{1}{b^5},v^5). \label{sg2}
 \end{eqnarray}
 For the interaction between a D0-D4 bound state and a D0-anti-D4 bound state
 we generalize the calculation in \cite{CT1},
 to find the potential:
 \begin{eqnarray}
   {\cal V} & \approx & 
  \frac{1}{4 b^3} \left[4 n_4 N_4 +
 (n_0 N_4 + N_0 n_4 )v^2 
 +  \frac{1}{4} ( n_0 N_0 + n_4 N_4 + 2 n_0 N_4+ 2 N_0 n_4  ) v^4 \right] \nonumber \\
& &
+O(\frac{1}{b^6},v^6). \label{sg3}
 \end{eqnarray}
The results agree with the matrix theory calculation at large $N_0$ and $n_0$.
 Note that the results for potentials (or corresponding phase
 shifts ($\d = \int dt \, {\cal V} (\sqrt{b^2+ (vt)^2})$)
 in matrix theory can also be compared directly to string
 theory calculations \cite{L2} \cite{L4}  .

\setcounter{equation}{0}
 \section{Technical details}
 Some of the technical details for determining the action
 (\ref{action}) are assembled here. We refer to \cite{KK} and
 \cite{Ja} for the rules to convert matrices into functions and traces
into integrals.
 We only keep  the relevant terms
and consider static configurations only.
 First we define the non-center-of-mass coordinates -- the center
 of mass coordinates just describe overall movements of the system in which
 we are not interested --:
 \begin{eqnarray}
 u_i &=& \frac{x_i+y_i}{2}.
 \end{eqnarray}
 Using the definitions given in the body of the text (\ref{def}), we can
 write down the commutators of the coordinate fields relevant to the
 problem in a reasonably compact form\footnote{We leave out the  factors
 of the zero brane density $\r_0$  to avoid cluttering the formulas
 even more. They can easily be added in \cite{KK} \cite{Ja}}: 

\begin{eqnarray}
   \left[ X^1,X^2 \right] =  & \qquad  & \hspace{10 cm}
\end{eqnarray}
\vspace{- 0.9 cm}
{\fontsize{8}{1}
\begin{eqnarray}
  & & \left (
 \begin{array}{cccc} i c^2 f^{(1)}_{x_1
 x_2} -i c + i  |\f|^2 & 0 &
 -\frac{c}{\sqrt{2}} (\nabla^{(1,-3)}_{u_2}+ i
 \nabla^{(1,-3)}_{u_1}) \f  & 0  \\  \ast  & i c^2 f^{(2)}_{x_1
 x_2} +i c -i |\f'|^2  & 0 &
 -\frac{c}{\sqrt{2}}(-\nabla^{(2,-4)}_{u_2}+ i
 \nabla^{(2,-4)}_{u_1} \f') \\ \ast & \ast &
 i c^2 f^{(3)}_{y_1 y_2} +i c -
 i |\f|^2  & 0
 \\ \ast & \ast & \ast & i
 c^2 f^{(4)}_{y_1 y_2}-i c + i |\f'|^2
         \end{array} \right)
 \nonumber 
\end{eqnarray}
}
\vspace{-0.8 cm}
\begin{eqnarray}
  \left[ X^3,X^4 \right] = & \qquad & \hspace{10 cm}
\label{com}  
\end{eqnarray}
\vspace{- 0.9 cm}
{\fontsize{8}{1}
\begin{eqnarray}
& & \left (
 \begin{array}{cccc} i c^2 f^{(1)}_{x_3
 x_4} -i c + i |\chi|^2 & 0 & 0 &
 -\frac{c}{\sqrt{2}}(\nabla^{(1,-4)}_{u_4}+ i
 \nabla^{(1,-4)}_{u_3}) \chi  \\ \ast & i c^2 
f^{(2)}_{x_3
 x_4} +i c - i |\chi '|^2 &
 -\frac{c}{\sqrt{2}} (-\nabla^{(2,-3)}_{u_4}+ i
 \nabla^{(2,-3)}_{u_3}) \chi '  & 0 \\ \ast & \ast
 & i c^2 f^{(3)}_{y_3 y_4}
  -i c + i |\chi '|^2 & 0 \\
 \ast & \ast & \ast &
 c^2 f^{(4)}_{y_3 y_4} +i c - i |\chi |^2
         \end{array} \right)  \nonumber 
\end{eqnarray} }
The other relevant commutators are all analogous to the following one:
\begin{eqnarray}
 \left[ X^1,X^3 \right] &=& \left (
 \begin{array}{cccc} i c^2 f^{(1)}_{u_1 u_3}   & -\frac{1}{2} (\f
 {\chi '}^{\ast}  - \chi {\f '}^{ \ast}  ) & -\frac{c}{\sqrt{2}}
 \nabla_{u_3}^{(1,-3)} \f & \frac{c}{\sqrt{2}}
 \nabla_{u_1}^{(1,-4)} \chi \\ \ast &  i c^2 f^{(2)}_{u_1 u_3}   &
 -\frac{c}{\sqrt{2}} \nabla_{u_1}^{(2,-3)} \chi ' &
 \frac{c}{\sqrt{2}} \nabla_{u_3}^{(2,-4)} \f '\\ \ast & \ast & i
 c^2 f^{(3)}_{u_1 u_3} &  \frac{1}{2} (\chi \f^{\ast}  - \f '
 {\chi '}^{\ast}  ) \\ \ast & \ast & \ast & i c^2 f^{(4)}_{u_1 u_3}
         \end{array} \right)  \nonumber \\
& & \label{com2}
\end{eqnarray}
The commutators are antihermitian.
 We then simplify the
 action by concentrating on the non-center-of-mass 
  fluctuations of the tachyon (compare \cite{Ja}):
 \begin{eqnarray}
 \phi (x_i,y_i) &=& 
 \phi(u_i) \sqrt{\d (x_1-y_1) \d(x_2-y_2) \d(x_3-y_3) \d(x_4-y_4 )}  \\
 \chi (x_i,y_i) &=& 
 \chi(u_i) \sqrt{\d (x_1-y_1) \d(x_2-y_2) \d(x_3-y_3) \d(x_4-y_4 )} 
 \end{eqnarray}
 Then the action reduces to 
 (\ref{action}), the integration running over four variables only.

  \end{document}